\documentclass{article}
\usepackage{PRIMEarxiv}
\usepackage[utf8]{inputenc} 
\usepackage[T1]{fontenc}    
\usepackage{hyperref}       
\usepackage{url}            
\usepackage{booktabs}       
\usepackage{amsfonts}       
\usepackage{nicefrac}       
\usepackage{microtype}      
\usepackage{lipsum}
\usepackage{fancyhdr}       
\usepackage{graphicx}       
\graphicspath{{media/}}     

\title{PROPERTIES OF HOT QUARK MATTER WITH NEUTRINO CONFINEMENT
IN THE NJL MODEL
}

\author{
 G.S.Hajyan\thanks{ghajyan@ysu.am}~~~   and ~ G.B. Alaverdyan\thanks {galaverdyan@ysu.am}\\
    Yerevan State University. Armenia \\
 } 

\begin{document}
\maketitle

\begin{abstract}

{The thermodynamic characteristics of hot $\beta$-equilibrium three-flavor quark matter with neutrino confinement are studied in terms of the local SU(3) Nambu-Jona-Lasinio (NJL) model, which also accounts
for the ‘t Hooft interaction which leads to mixing of quark flavors. For different temperatures
$T \in [20 \div 100]$ MeV and baryon number densities $n_B \in [0 \div 1.8]$ fm$^{-3}$ the constituent quark masses, quark
condensates, and relative contributions of individual types of particles to the pressure and chemical
potentials of the constituent particles are determined. In order to determine the role of neutrinos in the
hot quark matter, the pressures and energies in states with and without neutrinos are compared.
}

\end{abstract}


\section{Introduction}

This paper is a continuation of our earlier paper \cite{1} devoted to a study of several thermodynamic characteristics of hot $\beta$-equilibrium electrically neutral three-flavor quark matter with neutrino confinement. In Ref. 1, we
concentrated our attention on deriving the equation of state, composition, and speed of sound in terms of the local
SU(3) Nambu-Jona-Lasinio (NJL) model. Here we present new results obtained in later studies. In particular, here for
different values of the temperature $T$, we present the dependences on the baryonic charge concentration $n_B$ of the
constituent masses of $u$, $d$, and $s$ quarks, the quark condensates $\sigma_u$ , $\sigma_d$, and $\sigma_s$, of the relative contributions of
individual types of particles to the pressure, the energy $E_1=\varepsilon/n_B$ per baryon, as well as the chemical potentials of individual leptons when neutrinos are confined. To clarify the role of neutrino confinement in hot strange quark
matter (HSQM) for different values of the temperature, we have found the pressure as a function of the energy density
$\varepsilon$ and temperature, $P=P(\varepsilon,~T )$, as well as of the baryon number density $n_B$ and temperature, $P=P(n_B,~T)$, with and
without the presence of neutrinos. In the numerical calculations the presence of all types of leptons and antileptons
was taken into account, except for tau-leptons, which are absent for the baryon number densities $n_B$ and temperatures
$T$ examined here. The phenomenon of neutrino oscillations and the vector interaction between quarks were not taken
into account.

In this paper we shall use the "natural" system of units in which  $\hbar=c=k_B=1$.

\section{Quark matter in the local SU(3) Nambu-Jona-Lasinio model}

The NJL model has often been used recently to describe quark matter\cite{2,3}. The NJL model was originally
proposed for explaining the origin of the mass of a nucleon accounting for the spontaneous breaking of chiral
symmetry. However, later, in the 1970’s, this model was reformulated for describing quark matter \cite{4,5}. This model
has successfully reproduced many features of quantum chromodynamics (QCD) \cite{6,7,8}.
The Lagrangian density in terms of the SU(3) NJL model is given by

\begin{eqnarray}
\label{eq1} 
{\cal L}_{NJL}= \overline{\psi}\left(i\gamma ^{\mu }
\partial _{\mu }-\hat{m}_0\right)\psi+G_S \sum_{a=0}^{8}\left[(\overline{\psi}\lambda_a\psi)^2+(\overline{\psi}i\gamma_5\lambda_a\psi)^2\right] \nonumber\\ 
-K \left\{ det_f \left(\overline{\psi}(1+\gamma_5)\psi\right)+det_f \left(\overline{\psi}(1-\gamma_5)\psi\right) \right\}.
\end{eqnarray}

Here $\psi$ is the Fermion quark spinor field ${\psi_f}^c$c with three flavors $f = u,~d,~s$ and three colors $c=r~,g,~b$. The first
term is the density of the Dirac Lagrangian of free quark fields with a mass matrix of current quarks
${\hat{m}_0}=$ diag $(m_{0u},~m_{0d},~m_{0s})$. The second corresponds to a chirally-symmetric four-quark interaction with a coupling
constant $G$, where $\lambda_a$ ($a = 1, 2, ..., 8$) are the Gell-Mann matrices and generators of the SU(3) group in flavor space, $\lambda_0=\sqrt{2/3}\hat I$ ( $\hat I$ is the unit 3 $\times 3$ matrix). The third term corresponds to the six-quark Kobayashi-Maskawa-’t Hooft interaction \cite{9}, which leads to destruction of the axial $U_A(1)$ symmetry.

In the mean-field approximation the gap equations for the constituent masses of the quarks $M_u,~M_d,$ and $M_s$ are given by

\begin{eqnarray}
\label{eq2} 
 M_u=m_{0u}-4\,G_S\,\sigma_u+2\,K\sigma_d~\sigma_s, \nonumber \\
 M_d=m_{0d}-4\,G_S\,\sigma_d+2\,K\sigma_s~\sigma_u,\,\\
 M_s=m_{0s}-4\,G_S\,\sigma_s+2\,K\sigma_u~\sigma_d.~  \nonumber
\end{eqnarray}

Here $\sigma_{f}$, $f = u,~ d,~ s$ are the so-called quark condensates, which are defined by the expression

\begin{equation}
\label{eq23} 
\sigma_f=-\frac{3M_f}{\pi^2}\int_{0}^\Lambda
dk~\frac{k^2}{E_f(k,M_f)}\left[ 1-\frac{1}{1+e^{\left( E_f(k,M_f)-\mu_f \right)/T}}- \frac{1}{1+e^{\left( E_f(k,M_f)+\mu_f \right)/T}}  \right],
\end{equation}

where $\Lambda$ is the ultraviolet cutoff momentum, for which a need arises in connection with the non-renormalizable NJL
model, $E_f(k,M_f)=\sqrt{k^2+M_f^2}$ is the energy, and $\mu_f$ is the chemical potential of the quasiparticle-quarks of the
flavor $f=u,~ d,~ s$.

Knowledge of the constituent masses and quark condensates makes it possible, for specified values of the baryon number density $n_B$ and temperature $T$, to determine the other thermodynamic parameters of electrically neutral $\beta$-equilibrium quark matter with a constituent composition of particles $u$, $d$, $s$, $e$, $\nu_e$ , $\mu$ , $\nu_\mu$ , and $\nu_\tau$. A detailed discussion of the formalism used in our model and details of the numerical calculations can be found in Refs. 1 and 10.

As in Ref. 1, and in this article, numerical calculations are shown for the following parameters of the NJL model: $m_{0u}=m_{0d}=5.5$ MeV, $m_{0s}=140.7$ MeV, $\Lambda= 602.3$ MeV, $G=1.835/\Lambda^2$ and $K= 12.36/\Lambda^5$ obtained in Ref. 6 for reproducing the values of the coupling constant of the pion, $f_\pi=92.4$ MeV, as well as the masses of the $\pi$, $K$, $\eta$
and $\eta'$ mesons, $m_\pi=135$ MeV, $m_K=497.7$ MeV, $m_\eta=514.8$ MeV, and $m_\eta'=960.8$ MeV, respectively.

For the values of the specific $e$-lepton charge, a value of $Y_{L_e}={n_{L_e}}/{n_B}={(n_e+n_{\nu_e})}/{n_B}=0.4$ has been used, for the $\mu$-lepton charge, $Y_{L_\mu}={n_{L_\mu}}/{n_B}={(n_\mu+n_{\nu_\mu})}/{n_B}=0$ and for the $\tau$-lepton charge, $Y_{L_\tau}={n_{L_\tau}}/{n_B}={(n_\tau+n_{\nu_\tau})}/{n_B}=0$. In the density and temperature ranges we have examined, the conditions for creation of tau-leptons are not satisfied.

\section{Dependence of the characteristics of HSQM on the baryon number density at different temperatures}

The strong interaction among quarks and the presence of quark condensates lead to a change in the constituent mass of the quarks. Only for very high densities, when asymptotic freedom of quarks sets in, do the masses of the quarks $M_u$, $M_d$, and $M_s$ become equal to their current values $m_{0u}$,  $m_{0d}$, and $m_{0s}$.

\begin{figure}
\centering
\includegraphics[width=12 cm]{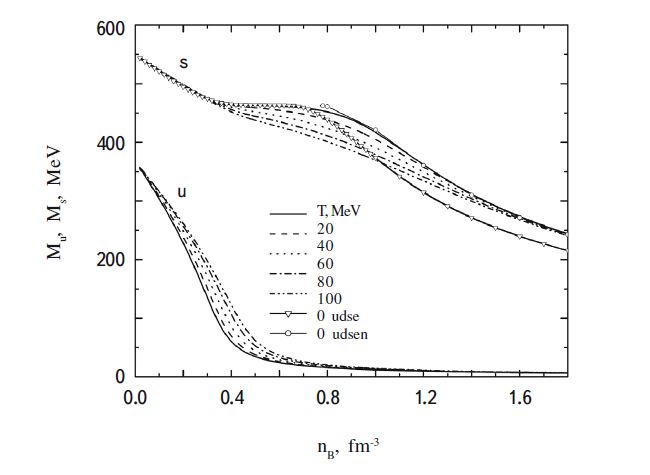}
\caption{The constituent masses of $u$ and $s$ quarks, $M_u$ and $M_s$, in HSQM with neutrino confinement as functions of the baryon number density $n_B$ for different temperatures $T$. The curve $udse$ corresponds cold SQM that is transparent for
neutrinos, and the curve $udse\nu$, to cold SQM with neutrino confinement.}
\end{figure}

Figure 1 shows the dependence of the masses of $u$ and $s$ quarks on the baryon number density $n_B$ for different temperatures $T$.

As Fig. 1 shows, with rising temperature and a constant baryon number density $n_B$ the mass of a $u$ quark increases. For high $n_B$ the state of the quarks approaches an asymptotically free state and, as expected, this change is insignificant. The situation is the same for very small $n_B$. For $n_B\approx 0.2 \div 0.6$ fm$^{-3}$, with temperature increases to $T = 100$ MeV the mass of a $u$ quark $M_u$
 increases by a factor of up to two. The mass of an $s$ quark increases
insignificantly with rising temperature for densities $n_B <0.3$ fm$^{-3}$. For $n_B >0.3$ fm$^{-3}$, the mass of an $s$ quark decreases with rising temperature.

This same figure shows the dependences of the mass of an $s$ quark on the baryonic charge concentration $n_B$ for $T = 0$, when there are no neutrinos (the curve $T = 0$, $udse$) and when neutrinos are present (the curve $T=0$, u$dse\nu_e$ ). The state $T = 0$, $udse\nu_e$ cannot be realized, since fully degenerate matter cannot confine neutrinos. This curve is introduced as the limit of the “hot” curves as $T \to 0 $.

Independently of the temperature, with increasing baryonic charge the masses of the constituent quarks approach their current values.

\begin{figure}
\centering
\includegraphics[width=12 cm]{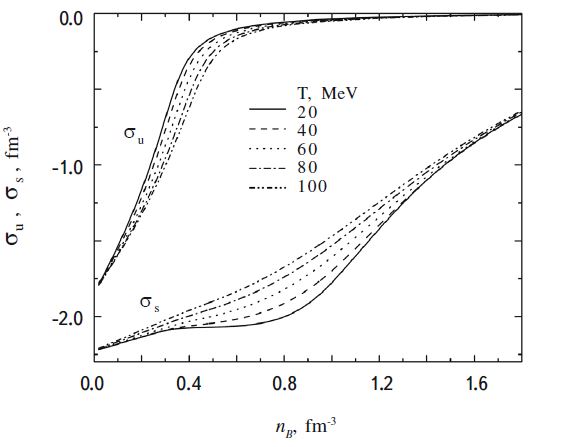}
\caption{The parameters of the quark condensates $\sigma_u$ and $\sigma_s$ as functions of the baryon number density $n_B$ for different temperatures $T$.}
\end{figure}
Figure 2 shows the quark condensates $\sigma_u$ and $\sigma_s$ as functions of the baryon number density $n_B$ for different values of the temperature $T$ and leptonic charges $Y_{L_e}= 0.4$, $Y_{L_\mu}= 0$, and $Y_{L_\tau}= 0$.

The higher the temperature of an HSQM, the lower the quark condensate $\sigma_u$. But for $\sigma_s$, on the other hand, the higher the temperature of an HSQM, the higher the value of the quark condensate $\sigma_s$.

The opposite character of the change in the quark masses in HSQM with rising temperature (Fig. 1) is caused precisely by the different character of the temperature dependence of the quark condensates $\sigma_u$ and $\sigma_d$ and of the quark condensate $\sigma_s$.

With increasing baryonic charge concentration $n_B$, the quark condensates vanish, independently of the temperature. Thus, the constituent masses of the quarks then tend to their current values, which is caused by the approach of the state of the HSQM to asymptotic freedom.

The dependences of the $u$ and $d$ quark masses, and the quark condensates $\sigma_u$ and $\sigma_d$ on the baryon number density $n_B$ for different temperatures are close; thus, here we show the results just for the $u$ quarks (see Figs. 1 and 2).

\begin{figure}
\centering
\includegraphics[width=12 cm]{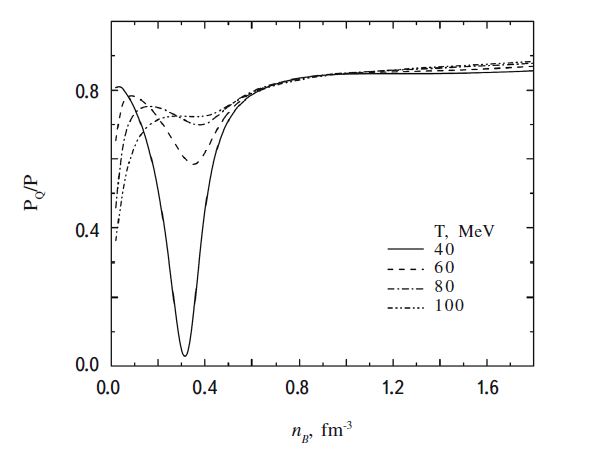}
\caption{The ratio of the quark pressure $P_Q$ to the total pressure $P$ in HSQM as a function of the baryon number density $n_B$ for different temperatures $T$.}
\end{figure}

Figure 3 shows the dependence of the ratio $P_Q/ P$ of the partial pressure of the quarks to the total pressure of HSQM for different temperatures $T$ on the baryon number density $n_B$.

The minimum of the curves at $n_B\approx 0.3$ fm$^{-3}$ is caused by the character of the quark interaction. The existence of this minimum is easily explained if the pressure of the quarks is represented as the sum of the pressure of the quarks at zero temperature $P_{Q0}$ and thermal corrections $\Delta P_{QT}$ to it. In the region of $n_B\approx 0.38$ fm$^{-3}$ according to Fig. 1 from Ref. 1, $P_{Q0}<0$ with a minimum at $n_B\approx 0.38$ fm$^{-3}$. The combined pressure $P_Q= P_{Q0}+\Delta P_{QT}$ above a temperature of $T\approx 20$ MeV for all $n_B$ is positive, but the negative minimum $P_{Q0}$ causes in the same region $n_B\approx 0.3$ fm$^{-3}$ the appearance of a positive minimum $P_Q$. With increasing temperature, the thermal correction increases so much that the minimum
becomes less marked.

For temperatures $T={40; 60}$ MeV, Figure 4 shows the dependences of the partial quark pressures $P_Q/P$ of the charged leptons (electrons and muons), $P_{e\mu}/P= (P_e+P_\mu) /P$, as well as of the corresponding neutrinos $P_{\nu}/P= (P_{\nu_e}+P_{\nu_\mu} )/P$. We note that the thermodynamic characteristics for this type of particle include contributions from both the particles themselves and the corresponding antiparticles.

Although the pressure of the leptons increases with rising temperature for a fixed concentration, their relative contribution to the pressure of the HSQM then decreases. For $T = const$ and $n_B > 0.8$ fm$^{-3}$, with increasing baryon number density, as opposed to the massive leptons, the relative contribution of the neutrinos to the pressure increases.

\begin{figure}
\centering
\includegraphics[width=16 cm]{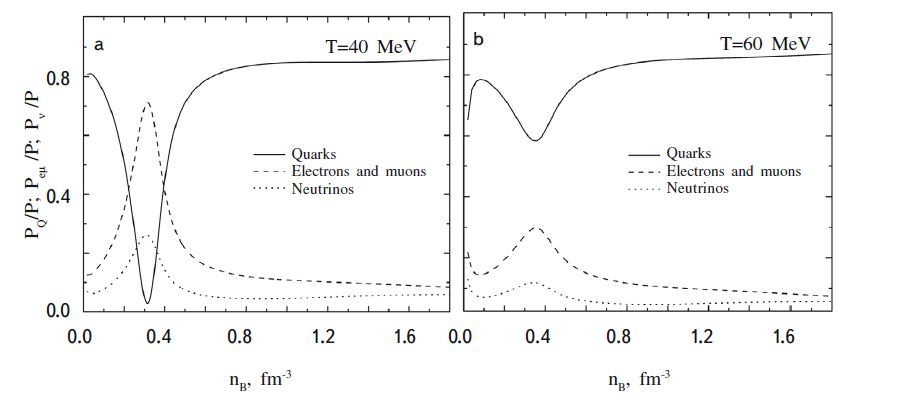}
\caption{Relative partial pressures of quarks $P_Q/P$, of charged leptons (electrons and
muons) $P_{e\mu} / P =(P_e + P_\mu )/ P$, and the same for the corresponding neutrinos,
$P_\nu / P =(P_{\nu_e} + P_{\nu_\mu} )/ P$ as functions of the baryon number density $n_B$ for $T = \{40;100\}$ MeV.}
\end{figure}

It should be noted that as $n_B\to 0$ in HSQM the quark pressure tends to zero, while the relative partial pressures of the leptons are nonzero, with $(P_e+P_\mu) /P\to 1$, since even for zero lepton charge in a thermodynamic equilibrium, infinitely rarefied, sufficiently hot medium, an electron-positron gas and thermal radiation exist with a
total pressure proportional to the fourth power of the temperature [11]. This occurs in a very narrow region near $n_B=0$, which is not shown in Fig. 4. We note also that the pressures of the radiation and the tau-neutrino gas in an HSQM in the region of the densities of interest for the physics of superdense celestial objects can be neglected \cite{1}. Speaking of neutrino confinement in a rarefied HSQM is not reasonable. This kind of state of matter is not realized in nature.

The dependence of the chemical potentials of the individual leptons on the baryon number density $n_B$ for $T = \{20;100\}$ MeV is shown in Fig. 5.

\begin{figure}
\centering
\includegraphics[width=16 cm]{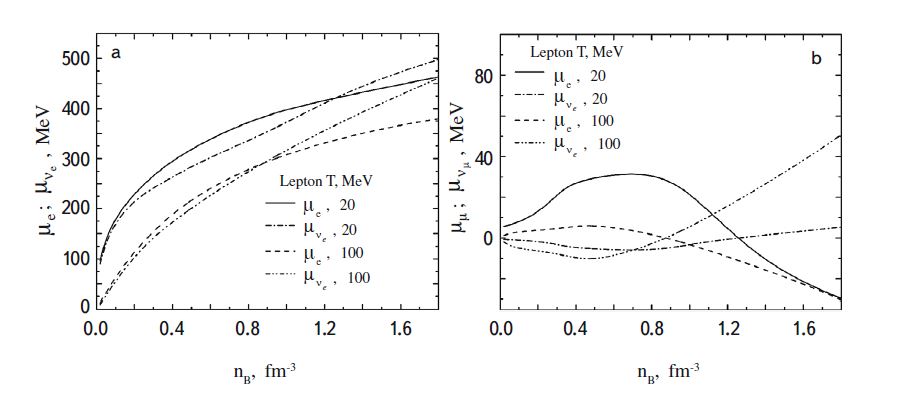}
\caption{The chemical potentials of individual leptons as functions of the baryon
number density $n_B$ for different temperatures $T = \{20;100\}$ MeV.}
\end{figure}

In the range of densities of interest for the astrophysics of superdense celestial objects ( $n_B>0.7$ fm$^{-3}$), the chemical potential $\mu_e$ of the electrons exceeds 300 MeV. Thus, the electron concentration is considerably higher than that of the positrons. The concentration of the muons is greater than that of the antimuons up to densities $n_B\approx 1.25$ fm$^{-3}$ for a temperature of $T = 20$ MeV and $n_B=0.87$  fm$^{-3}$ for a temperature of $T = 100$ MeV. Above these values of $n_B$, the chemical potential of the muons becomes negative and falls off rapidly. In this region the
concentration of the antimuons is much greater than the muon concentration.

When there are no neutrinos in the ordinary baryonic matter the chemical potentials of the electrons and muons do not exceed the rest energy of the $\pi$-meson \cite{12}, and, in HSQM, a few tens of MeV \cite{13,14}. Neutrino confinement in HSQM ensures a high value of the leptonic charge, which also leads to such large values of the chemical potentials of the leptons. As shown in Fig. 6, the chemical potential of the electron neutrinos $\mu_{\nu_e}$ reaches values of 450 MeV. Thus, the concentration of electron neutrinos is much higher than the concentration of positron neutrinos.

For a fixed value of $T$, when the chemical potential of a given lepton is higher, its pressure will be higher. Thus, the partial pressure of the leptons is mainly determined by the electrons and the electron neutrinos. Although the chemical potential of the electrons in HSQM for $n_B> 1.3$ fm$^{-3}$ is less than the chemical potential of the electron neutrinos, for an electron degeneracy factor that is twice that of the neutrinos, their pressure turns out to be higher than that of the neutrinos. The increase in the relative neutrino pressure in the range of baryon number density
$n_B> 0.8$ fm$^{-3}$ (Fig. 4), is caused by a large increase in the chemical potentials of the neutrinos (Fig. 5).

\section{The role of neutrino confinement in HSQM}

The presence of neutrinos in HSQM in a thermodynamic equilibrium state with the matter leads to certain changes in the characteristics of the HSQM. In this paper we compare those characteristics of the HSQM which are of great interest for the astrophysics of superdense celestial objects. 

Figure 6 shows the energy of HSQM for $T = \{20;100\}$ MeV per baryonB $E_1=\varepsilon /n_B$ as a function of the baryon number density $n_B$ with (smooth curves) and without (dashed curves) neutrino confinement.

\begin{figure}
\centering
\includegraphics[width=14 cm]{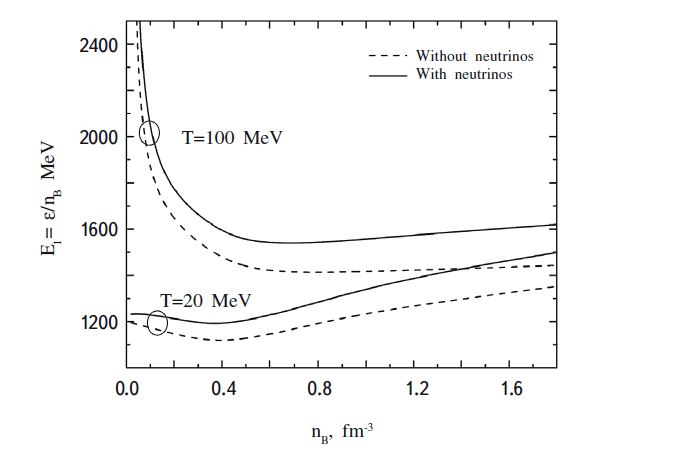}
\caption{The energies of HSQM per unit baryon charge $E_1$ as functions of baryon number density $n_B$ with (smooth curves) and without (dashed curves) neutrino confinement for temperatures $T = \{20;100\} $ MeV.}
\end{figure}

It is clear that neutrino confinement, which ensures a high relative lepton charge $Y\equiv Y_{L_e}\equiv 0.4$, increases $E_1$ to ten percent, which, in turn, leads to an increase in the pressure for a given baryon number density $n_B$. This is a result of the presence of a substantial number of leptons and the change in the composite composition of the quarks (see Fig. 5 in Ref. 1).

As opposed to HSQM without neutrino confinement, in which the relative number of electrons is negligibly small, with neutrino confinement their number reaches 10-15 percent of the number of quarks. Although the chemical potential of the electron neutrinos is greater than the chemical potential of the electrons, their number is less than the number of electrons because of the difference in the degeneracy multiplicities (see Eq. (9) in Ref. 1).

For the physics of superdense celestial objects the equation of state of superdense matter $P=P(\varepsilon,~ T)$ has a special significance. The Tolman-Oppenheimer-Volkoff (TOV) equations with the equations for energy transport and energy production determine the internal structure and integral parameters of the star. If the star is isothermal and
radiates only mainly because of stored thermal energy then the TOV equations are sufficient for solving the problem.

Despite its imperfections, the MIT quark bag model, as opposed to the Nambu-Jona-Lasinio model offers the possibility of determining the thermodynamic characteristics and equation of state of HSQM with neutrino confinement without cumbersome calculations \cite{15}. Based on the MIT model, in Ref. 15 it is shown that the pressure of
HSQM as a function $P=P(\varepsilon,~ T)$ of the energy density $\varepsilon$ and temperature for $\varepsilon=const$ depends weakly on temperature. In fact, the denser and hotter the HSQM is, the weaker this dependence will be.

\begin{figure}
\centering
\includegraphics[width=12 cm]{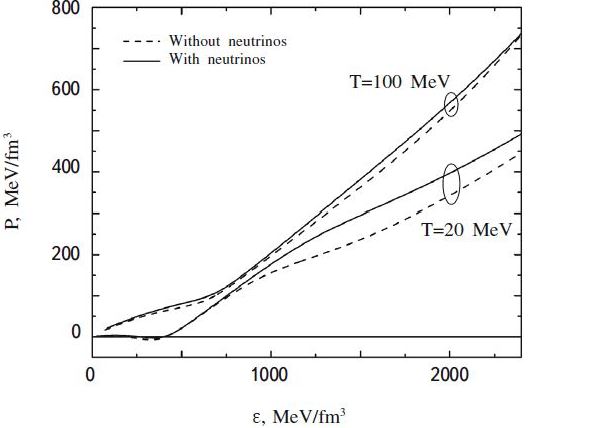}
\caption{The pressure $P$ of HSQM as a function of energy density $\varepsilon$ with (smooth curves) and without (dashed curves) neutrino confinement for temperatures $T = \{20; 100\}$ MeV.}
\end{figure}

Figure 7 shows the dependences of the pressure of HSQM on H according to the NJL model for temperatures of 20 and 100 MeV with (smooth curves) and without (dashed curves) neutrino confinement.

As opposed to the MIT model, in the NJL model the pressure of HSQM as a function of the energy density $\varepsilon$ and temperature $T$ for a fixed value of $\varepsilon$, depends strongly on temperature, regardless of neutrino confinement.
However, while for $T = 20$ MeV the pressures of HSQM with and without the presence of neutrinos differ by ~25 $\%$ with rising temperature this difference decreases, and for $T = 100$ MeV it falls below 5 $\%$.

Figure 8 shows the dependences of the pressure of HSQM on the baryon number density $n_B$ according to the NJL model with (smooth curves) and without (dashed curves) neutrino confinement for temperatures of 20 and 100 MeV. It is clear that in both cases the temperature dependence of the pressure is fairly strong.

\begin{figure}
\centering
\includegraphics[width=12 cm]{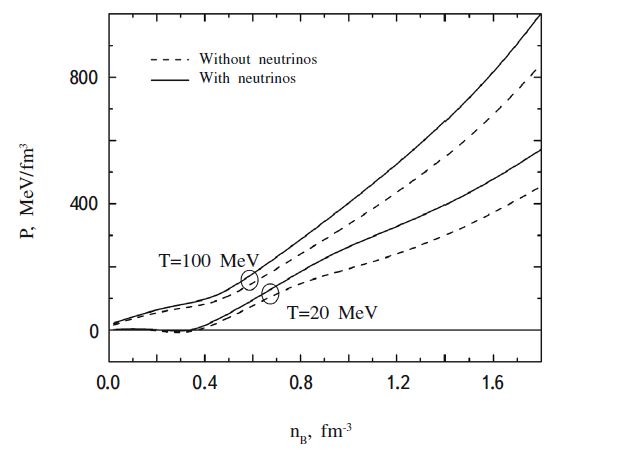}
\caption{The pressure $P$ of HSQM as a function of baryon number density $n_B$ with (smooth curves)
and without (dashed curves) neutrino confinement for temperatures $T = \{20; 100\}$ MeV.}
\end{figure}

It is also clear that, as opposed to the case of $P=P(\varepsilon,~ T)$, in the case of $P=P(n_B,~ T)$ for a fixed value of $n_B$, as the temperature rises the difference between the pressures with and without neutrino confinement does not decrease.

This means that in a proto-neutron star the radius and energy density distribution in a very hot ($T\sim 100$ MeV) quark core with a total baryonic charge $N_B$ will be almost independent of the presence of neutrinos in it. We recall that this is caused by the fact that only the energy density and pressure show up in the TOV equations.

\section{ Conclusion and critical comments}

The thermodynamic characteristics of hot quark matter which is opaque to neutrinos have been determined in the framework of the NJL model without accounting for neutrino oscillations, and vector and axial-vector quark interaction channels. Three sorts of neutrinos are taken into account in our calculations in accordance with the standard model for the theory of elementary particles. The ratio of the combined lepton and baryon charges of HSQM is taken to be 0.4, which is close to the value of this parameter in the matter of a pre-supernova star. It is shown that the presence of leptons in HSQM in this amount basically leads to strong changes in the temperature dependence's of the thermodynamic characteristics, the parameters of the particles, and the equation of state of matter. The applicability of our results for the construction of models of the core of a proto-quark star depends on how well justified the choice of the relative lepton charge of 0.4 in HSQM is. Of course, during a supernova explosion to the
formation of a proto-neutron star with a quark core, neutrinos will partially carry away the leptonic charge of the dense core of the pre-supernova star. If $\beta$-equilibrium in the central regions of a proto-neutron star is established in matter that is already opaque to neutrinos, then the choice of this value for the relative lepton charge is justified. It is possible that this may take place only under certain conditions. The question of “under what conditions?” can be answered only by solving the detailed time dependent problem of a the explosion of a supernova star with different initial conditions.

\section*{Acknowledgments}
This work was carried out in the scientific-research laboratory for the physics of superdense stars in the department of applied electrodynamics and modeling at Yerevan State University, financed by the committee on science of the Ministry of Education, Science, Culture, and Sport of the Republic of Armenia.


\begin{thebibliography}{}
\bibitem{1}	G. S. Hajyan and G. B. Alaverdyan, {\em Astrophysics} {\bf 64}, 370 (2021).
\bibitem{2}  Y. Nambu and G. Jona-Lasinio, {\em  Phys. Rev.} {\bf 122}, 345 (1961).
\bibitem{3}	Y. Nambu and G. Jona-Lasinio, {\em  Phys. Rev.} {\bf 124}, 246 (1961).
\bibitem{4}	O. Eguchi, {\em Phys. Rev. D} {\bf 14}, 2755 (1976).
\bibitem{5} E. Kikkawa, {\em  Prog. Theor. Phys.} {\bf 56}, 947, 1976).
\bibitem{6}	P. Rehberg, S. P. Klevansky, and J. Hüfner, {\em  Phys. Rev. C} {\bf 53}, 410 
\bibitem{7}	M. Buballa, {\em Phys. Rep.} {\bf 407}, 205 (2005).
\bibitem{8}	M. K. Volkov and A. E. Radzhabov, {\em  Usp. Fiz. Nauk} {\bf 176}, 569 (2006). 
\bibitem{9}	G. ‘t Hooft, {\em  Phys. Rev. Lett.} {\bf 37}, 8 (1976).
\bibitem{10} G. Alaverdyan, {\em  Symmetry} {\bf 13}, 124 (2021).
\bibitem{11} Ya. B. Zel’dovich and I. D. Novikov, {\em Relativistic Astrophysics} [in Russian], Nauka, Moscow (1967).
\bibitem{12} V. A. Ambartsumyan and G. S. Saakyan, {\em  Voprosy kosmogonii}[in Russian] {\bf 9}, 91 (1963).
\bibitem{13} G. S. Hajyan and A. G. Alaverdyan, {\em  Astrophysics} {\bf 57}, 559 (2014).
\bibitem{14} Yu. L. Vartanyan, Sh. R. Melikyan, and A. A. Shaginyan, {\em Astrophysics} {\bf 55}, 429 (2012).
\bibitem{15} G. S. Hajyan, {\em  Astrophysics} {\bf 61}, 511 (2018).

\end{thebibliography}
\end{document}